\documentclass[titlepage,twoside,12pt]{article}
\usepackage{amssymb}
\usepackage{amsfonts}
\begin{document}

{\bf \Large An omission in Special Relativity} \\ \\

Elem\'{e}r E Rosinger \\
Department of Mathematics \\
and Applied Mathematics \\
University of Pretoria \\
Pretoria \\
0002 South Africa \\
eerosinger@hotmail.com \\ \\

{\bf Abstract} \\

In Special Relativity there are {\it infinitely} many equivalence classes of frames of
reference. In each such class every two frames of reference move uniformly with respect to one
another. \\
It appears that the following questions have never been asked : \\
In which of such equivalence classes of frames of reference is Special Relativity supposed to
hold ? \\
How can we identify such a class, and do so theoretically or empirically ? \\ \\

{\bf \large 1. Equivalence classes of frames of reference} \\

The two principles of Special Relativity, Einstein [2, p. 39], Durell [p. 30], are : \\

1.~ There is no absolute motion. Consequently, it is not possible to detect uniform motion.
Furthermore, in two frames of reference which move uniformly with respect to one another, the
laws of physics are the same.\\

2.~ In all forms of wave motion, the velocity of the wave propagation does not depend on the
velocity of the source. Consequently, the velocity of light propagation in vacuum does not
depend on the velocity of its source. \\

Let us denote by ${\cal K}$ the set of all space-time frames of reference given by
corresponding rigid solids and clocks. Each such frame of reference $K$ in ${\cal K}$ is
supposed to be associated with a respective coordinate system $(x, y, z, t)$. \\

Now, in view of the first above principle, let us define on the set ${\cal K}$ of space-time
frames of reference a binary relation $\approx$ as follows. For every $K, K^\prime \in
{\cal K}$ we have \\

$~~~~~~ K ~\approx~ K^\prime ~~~\Longleftrightarrow~~~ K^\prime
                  ~~\mbox{moves uniformly with respect to}~~ K $ \\

In terms of the Lorentz transformations, this means that the coordinate systems $(x, y, z, t)$
in $K$, and respectively $(x^\prime, y^\prime, z^\prime, t^\prime)$ in $K^\prime$, are related
according to the well known relations \\

$~~~~~~ \begin{array}{l}
                x^\prime ~=~ ( x - v t ) / \sqrt ( 1 - v^2 / c^2 ) \\
                y^\prime ~=~ y \\
                z^\prime ~=~ z \\
                t^\prime ~=~ ( t - v x / c^2 ) / \sqrt ( 1 - v^2 / c^2 )
        \end{array} $ \\

assuming that the frame of reference $K^\prime$ moves along the coordinate axes $x$ of $K$ and
does so with constant velocity $v$. \\

It is easy to see that, with the above definition, the binary relation $\approx$ is an {\it
equivalence relation} on ${\cal K}$, namely : \\

a)~ it is {\it reflexive}, that is, we have \\

$~~~~~~ K \approx K, ~~~\mbox{for all}~~ K \in {\cal K}$ \\

b)~ it is {\it symmetric}, that is, we have \\

$~~~~~~ K \approx K^\prime ~\Longrightarrow~
                  K^\prime \approx K, ~~~\mbox{for all}~~ K, K^\prime \in {\cal K}$ \\

c)~ it is {\it transitive}, that is, we have \\

$~~~~~~ K \approx K^\prime,~ K^\prime \approx K^{\prime \prime} ~\Longrightarrow~
      K \approx K^{\prime \prime}, ~~~\mbox{for all}~~
                  K, K^\prime, K^{\prime \prime} \in {\cal K}$ \\

In this way, we can define the {\it quotient} set \\

$~~~~~~ {\cal E K} ~=~ {\cal K} / \approx $ \\

whose elements are the {\it equivalence classes} \\

$~~~~~~ K_\approx ~=~ \{~ K^\prime \in {\cal K} ~|~ K^\prime \approx K ~\} $ \\

of frames of reference. Thus each such equivalence class is made up of all the frames of
reference which move uniformly with respect to one another. Consequently, each such class
$K_\approx$ contains infinitely many frames of reference, each two of them being equivalent. \\ \\

{\bf \large 2. Too many equivalence classes : two questions} \\

The important point to note is that in ${\cal E K}$ itself there are {\it infinitely} many
{\it different} equivalence classes $K_\approx$ of frames of reference. \\
Indeed, the Earth and the Moon, for instance, obviously belong to two different such classes,
since the Earth and the Moon do not move uniformly with respect to one another. And more near
to us, the Earth and any car which moves along a curved part of a road also belong to two
such different classes. \\

Now clearly, the laws of motion {\it cannot} be identical in all these infinitely many
different equivalence classes $K_\approx$ in ${\cal E K}$. For instance, they cannot be
identical in the equivalence class of the Earth and in the equivalence class of a car moving
on Earth along a curved part of the road. \\

And then the {\it question} arises :

\begin{itemize}

\item In which of the equivalence classes $K_\approx$ of frames of reference in ${\cal E K}$
is Special Relativity supposed to hold, yielding in particular the canonical forms of the laws
of motion ?

\end{itemize}

Furthermore, in case there exists such an equivalence class of frames of reference $K_\approx
\in {\cal E K}$, then :

\begin{itemize}

\item How can we identify it, and do so theoretically or empirically ?

\end{itemize}

It appears that, so far, there has not been an explicit enough awareness of the fact that

\begin{itemize}

\item ${\cal E K}$ does contain {\it infinitely} many different equivalence classes of frames
of reference

\item the laws of motion {\it cannot} be identical in all of these infinitely many frames of
reference.

\end{itemize}

Consequently, the above questions appear not to have been considered. \\ \\

{\bf \large 3. The case of Newtonian mechanics} \\

In Newtonian mechanics, unlike in Special Relativity, space and time are considered
separately. \\

As far as time is concerned, it is simply considered to be absolute, and it is assumed to flow
uniformly from past into the future, regardless of whatever may happen in space. Therefore,
there is no attempt whatsoever to further elaborate on its possible nature, and it is simply
given by a unique universal frame of reference $T$, with its coordinate $t$. \\

Concerning space, the various frames of reference $L^{Newton}$ are given by rigid bodies and
are endowed with respective coordinate systems $( x, y, z )$. \\

It follows that space-time in Newtonian mechanics is represented by frames of reference
$K^{Newton} = L^{Newton} \times T$, with the respective coordinate systems $( x, y, z, t )$.
In this way, in Newtonian mechanics, the equivalence between the coordinate systems
$(x, y, z, t)$ in $K^{Newton}$, and respectively $(x^\prime, y^\prime, z^\prime, t^\prime)$ in
$K^{\prime~Newton}$, is given according to the well known relations of Galilean relativity \\

$~~~~~~ \begin{array}{l}
                x^\prime ~=~ x - v t \\
                y^\prime ~=~ y \\
                z^\prime ~=~ z \\
                t^\prime ~=~ t
        \end{array} $ \\

assuming again that the frame of reference $K^{\prime~Newton}$ moves along the coordinate axes
$x$ of $K^{Newton}$ and does so with constant velocity $v$. \\

However, in view of this principle of Galilean relativity, which is the content of Newton's
first law, the issue arises to identify at least one such a spatial frame of reference which
is tied to absolute space. \\
Newton suggested for that purpose that the {\it faraway stars} may offer a good {\it
approximation} to such a spatial frame of reference which is tied to absolute space. Let us
denote by $K^{Faraway Stars}$ that frame of reference. \\

Now, if we denote by ${\cal K}^{Newton}$ the set of all Newtonian space frames of reference
$K^{Newton}$, then similar with section 1, this time however using the Galilean relativity
instead of the Lorentzian one, we can define on ${\cal K}^{Newton}$ a corresponding
equivalence relation $\approx$, and obtain the set \\

$~~~~~~ {\cal E K}^{Newton} ~=~ {\cal K}^{Newton} / \approx $ \\

of equivalence classes \\

$~~~~~~ K^{Newton}_\approx ~=~ \{~ K^{\prime~Newton} \in {\cal K}^{Newton} ~|~
                                       K^{\prime~Newton} \approx K^{Newton} ~\} $ \\

Clearly, the set ${\cal E K}^{Newton}$ is again infinite. In other words, there are infinitely
many different equivalence classes $K^{Newton}_\approx$. \\

In this case, however, the only such equivalence class we are interested in when considering
Newton's first law is $K^{Faraway Stars}_\approx$. Thus the above two questions - which arise
in Special Relativity - do {\it not} arise in Newtonian mechanics, since from the start they
are answered implicitly by the fundamental assumptions about absolute space and absolute
time. \\ \\

\newpage

{\bf \large 4. General Relativity answers the two questions} \\

In General Relativity the concept of {\it equivalent} frames of reference is much {\it
enlarged}. Indeed, instead of the restrictive Galilean or Lorentzian conditions on equivalence,
now two frames of reference are considered equivalent if there is a ${\cal C}^2$-smooth
diffeomorphism between their respective space-time coordinate systems. Consequently, for all
purposes, there is only {\it one single} class of equivalent frames of reference. Thus the
above two question once again {\it cannot} arise from the very beginning. \\ \\

{\bf \large 5. Conclusions} \\

It follows that, apart from all other possible reasons, one of the most important
considerations for going from Special Relativity to General Relativity is to introduce an
equivalence relation between frames of reference which leads to {\it one single} corresponding
equivalence class. In this way one can, among others, go beyond the two questions in section
2. \\

Galilean and Lorentzian relativity, and the corresponding equivalence between frames of
reference, happened to be {\it fixated} upon a {\it first relativistic realization}, namely,
that the laws of motion do not depend on the constant velocity of the observer. Consequently,
the laws of motion have to be given by differential equations which are of {\it second order}
in the coordinates of the moving objects. Aristotle was not aware of this first relativistic
phenomenon. Instead, he considered that the velocity of a moving object is proportional to the
force applied upon it, which in terms of differential equations would only lead to a first
order one in the coordinates of the object. Such a first order equation would, of course, be
in an obvious contradiction with the fact that one can set up {\it two} arbitrary and
independent {\it initial} conditions, when establishing the motion of a material point. \\

As seen in section 2, the first relativistic realization of Galileo, Lorentz and of Special
Relativity leads to {\it infinitely many} equivalence classes of frames of reference. This is
not a problem in Newtonian mechanics where one assumes absolute space and absolute time.
However, it becomes a problem in Special Relativity, precisely because of the {\it lack} of
absolute space or absolute time which is now assumed. \\

And it needs a {\it second relativistic realization}, as provided by General Relativity, in
order to abolish any and all distinctions between frames of reference, distinctions which may
be relevant with respect to the laws of motion. This second relativistic realization is that
the laws of motion have to be {\it covariant} with respect to arbitrary ${\cal C}^2$-smooth
diffeomorphic transformations of coordinates. Consequently, all frames of reference are now
equivalent, and we only have one equivalence class of frames of reference. \\

In this way, from the earlier Newtonian absoluteness of space-time, we have moved in General
Relativity to what may be seen as the {\it absoluteness of the unique class equivalent frames
of reference}. \\

\end{document}